\documentclass[twocolumn,showpacs,preprintnumbers,floatfix,nofootinbib]{revtex4}
\usepackage{amssymb}
\usepackage{amsmath}
\usepackage{graphicx}
\usepackage{dcolumn}
\usepackage{bm}
\usepackage[subfigure]{graphfig}
\usepackage{makecell}
\usepackage{color}

\begin{document}

\title{$\Lambda _{c}^{+}(2940)$ photoproduction off the neutron}
\author{Xiao-Yun Wang$^{1,2,3}$}
\thanks{xywang@impcas.ac.cn}
\author{Alexey Guskov$^{4}$}
\thanks{avg@jinr.ru}
\author{Xu-Rong Chen$^{1,3}$}
\affiliation{$^1$Institute of Modern Physics, Chinese Academy of Sciences, Lanzhou
730000, China\\
$^2$University of Chinese Academy of Sciences, Beijing 100049, China\\
$^3$Research Center for Hadron and CSR Physics, Institute of Modern Physics
of CAS and Lanzhou University, Lanzhou 730000, China\\
$^4$Joint Institute for Nuclear Research, Dubna 141980, Russia}

\begin{abstract}
By assuming that the $\Lambda _{c}^{\ast }(2940)$ is a $pD^{\ast 0}$
molecular state with spin-parity $J^{P}=\frac{1}{2}^{+}$ and $J^{P}=\frac{1}{%
2}^{-}$, the photoproduction of charmed $\Lambda _{c}^{\ast }(2940)$ baryon
in the $\gamma n\rightarrow D^{-}\Lambda _{c}^{\ast }(2940)^{+}$ process is
investigated with an effective Lagrangian approach. It is found that the
contributions from $t$-channel with $D^{\ast }$ exchange are dominant, while
the $s$-channel with nucleon pole exchange give a sizeable contribution
around the threshold. The contributions from the $u$-channel and contact
term are very small. The total cross section of the $\gamma n\rightarrow
D^{-}\Lambda _{c}^{\ast }(2940)^{+}$ reaction is estimated, which indicate
it is feasible to searching for the charmed $\Lambda _{c}^{\ast }(2940)$
baryon at the COMPASS experiment.
\end{abstract}

\pacs{14.20.Lq, 13.60.Rj, 13.30.Eg}
\maketitle

\section{Introduction}

Searching and explaining the exotic states which may consist of the non $q%
\bar{q}$ and $qqq$ configurations, have becoming a very interesting topic in
hadron physics. Actually, the structure of baryon is more intriguing than
that of the meson. Recently, some charmed baryons have been experimentally
identified \cite{pdg,ek10}, which provide an ideal place to investigate the
dynamics of the light quarks in the environment of a heavy quark. For
example, the charmed baryon $\Lambda _{c}^{\ast }(2940)$ has aroused
intensive studies on its nature.

The charmed baryon $\Lambda _{c}^{\ast }(2940)$ was first announced by the
\textit{BABAR} Collaboration \cite{babar07} by analyzing the $pD^{0}$
invariant mass spectrum. Later, the Belle Collaboration \cite{belle07}
confirmed it as a resonant structure in the final state $\Sigma
_{c}(2455)^{0,++}\pi ^{\pm }\rightarrow \Lambda _{c}^{+}\pi ^{+}\pi ^{-}$.
The values for the mass and width of the $\Lambda _{c}^{\ast }(2940)$ state
were reported by both Collaborations \cite{babar07,belle07}, which are
consistent with each other:%
\begin{eqnarray*}
\text{\textit{BABAR }}\text{: } &M&=2939.8\pm 1.3\pm 1.0\text{ MeV,} \\
&\Gamma& =17.5\pm 5.2\pm 5.9\text{ MeV,} \\
\text{Belle}\text{: }&M& =2938.0\pm 1.3_{-4.0}^{+2.0}\text{ MeV,} \\
\text{ \ \ \ \ } &\Gamma& =13_{-5-7}^{+8+27}\text{ MeV.}
\end{eqnarray*}

However, the spin-parity of the $\Lambda _{c}^{\ast }(2940)$ state have
still not been determined in experiment. Different theoretical groups \cite%
{dong14,dong10,dong101,xg07,sc81,cc07,de08,xh08,hy07,sm08,wr08,cg09,dg11,he10,pg13}
have performed theoretical studies of $\Lambda _{c}^{\ast }(2940)$ by
assuming different assignment for its spin-parity $J^{P}=\frac{1}{2}^{\pm },%
\frac{3}{2}^{\pm },\frac{5}{2}^{\pm }$. For example, by assuming the $%
\Lambda _{c}^{\ast }(2940)$ as a $pD^{\ast 0}$ molecular state, the
spin-parity of $\Lambda _{c}^{\ast }(2940)$ was assigned to be $\frac{1}{2}%
^{\pm }$ in Refs. \cite{dong10,dong101,he10}. Besides supposing $\Lambda
_{c}^{\ast }(2940)$ to be a hadronic molecular state, the $\Lambda
_{c}^{\ast }(2940)$ also is explained as a conventional charmed baryon \cite%
{sc81} with $J^{P}=\frac{3}{2}^{+}$ or $J^{P}=\frac{5}{2}^{-}$. Since the
the nature of $\Lambda _{c}^{\ast }(2940)$ is still unclear, more work is
needed to determine its real inner structure.

Until now, all experimental observations of $\Lambda _{c}^{\ast }(2940)$
have been from the $e^{+}e^{-}$ collision \cite{babar07,belle07}. Thus it is
interesting to study the production of $\Lambda _{c}^{\ast }(2940)$ in other
process. In Refs. \cite{dong14,he11}, the production of $\Lambda _{c}^{\ast
}(2940)$ by $\bar{p}p$ annihilation are proposed, while the production of $%
\Lambda _{c}^{\ast }(2940)$ via $\pi $ meson induced nucleon is discussed in
Ref. \cite{xie15}. However, one notice that there is no any relevant
informations about the photoproduction of $\Lambda _{c}^{\ast }(2940)$. Thus
the studies on the photoproduction of $\Lambda _{c}^{\ast }(2940)$ are
highly necessary.

In this work, with an effective Lagrangian approach, the photoproduction of $%
\Lambda _{c}^{\ast }(2940)$ in the $\gamma n\rightarrow D^{-}\Lambda
_{c}^{\ast }(2940)^{+}$ process is investigated. Moreover, the feasibility
of searching for the charmed $\Lambda _{c}^{\ast }(2940)$ resonance is also
discussed. {It is shown that modern experiments based on energetic lepton
beams of high intensity like the COMPASS experiment at CERN \cite%
{Abbon:2007pq}\cite{compass} could be the promising platform for searching
for photoproduction of the charmed baryon $\Lambda _{c}^{\ast }(2940)$ and
study of its properties.}

This paper is organized as follows. After an Introduction, the formalism and
the main ingredients are presented. The numerical results and discussions
are given in Sec. III. In Sec. IV, the {$\Lambda _{c}^{\ast }(2940)$
production at COMPASS are discussed.} Finally, the paper ends with a brief
summary.

\section{Formalism}

In the present work, an effective Lagrangian approach in terms of hadrons is
adopted, which is an important theoretical method in investigating various
processes in the resonance region \cite%
{dong14,he11,xie15,zou03,xyw15,xy,epl15,prc15,epja15}.

\subsection{Feynman diagrams and effective Lagrangian densities}

Fig. 1 describes the basic tree level Feynman diagrams for the production of
$\Lambda _{c}^{\ast }(2940)$ ($\equiv \Lambda _{c}^{\ast }$) in $\gamma
n\rightarrow D^{-}\Lambda _{c}^{\ast +}$ reaction. These including the $t$%
-channel with $D^{+}$ and $D^{\ast +}$ exchange, $s$-channel with nucleon
pole exchange, $u$-channel with $\Lambda _{c}^{\ast }$ exchange and contact
term. Fig. 2 is the Feynman diagrams for the $\gamma n\rightarrow
D^{-}D^{0}p $ reaction.

\begin{figure}[tbph]
\begin{center}
\includegraphics[scale=0.45]{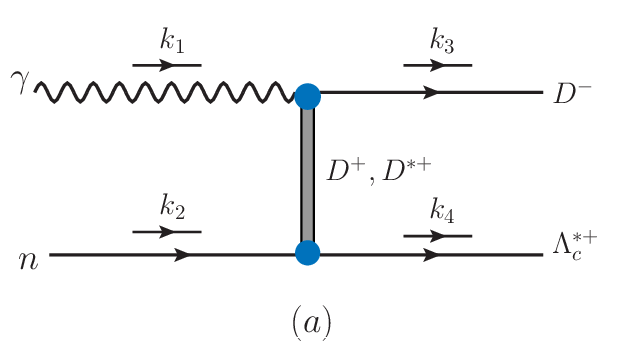} %
\includegraphics[scale=0.45]{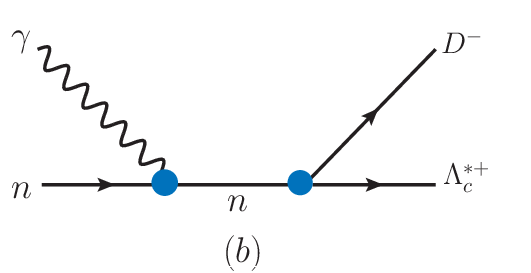} %
\includegraphics[scale=0.45]{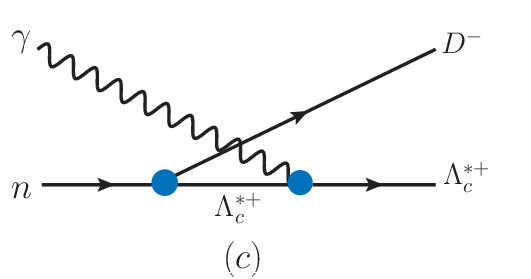} %
\includegraphics[scale=0.45]{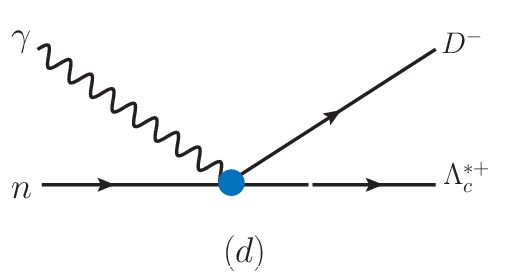}
\end{center}
\caption{(Color online) Feynman diagrams for the $\protect\gamma %
n\rightarrow D^{-}\Lambda _{c}^{\ast +}$ reaction. (a) $t$-channel; (b) $s$%
-channel; (c) $u$-channel; (d) contact term.}
\label{Fig:fyd}
\end{figure}
\begin{figure*}[tbph]
\begin{center}
\includegraphics[scale=0.4]{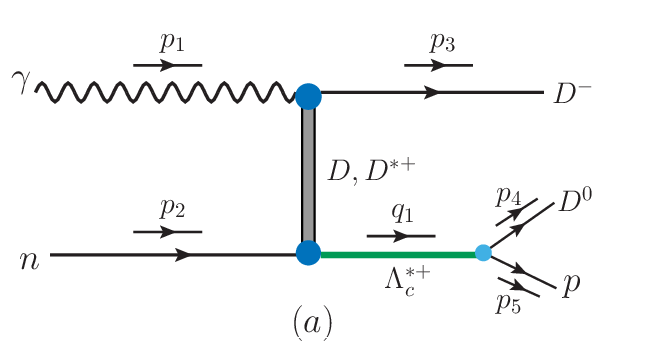} \includegraphics[scale=0.4]{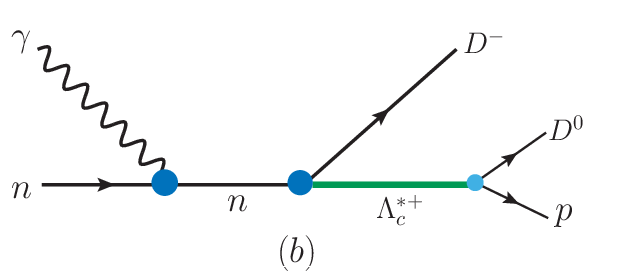}
\includegraphics[scale=0.4]{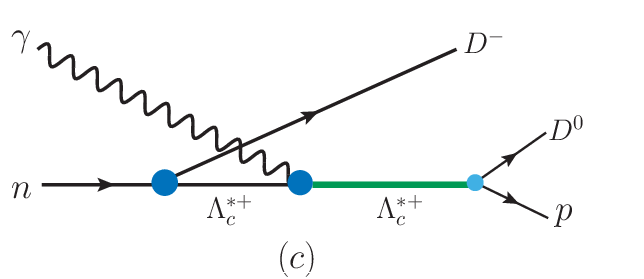} \includegraphics[scale=0.4]{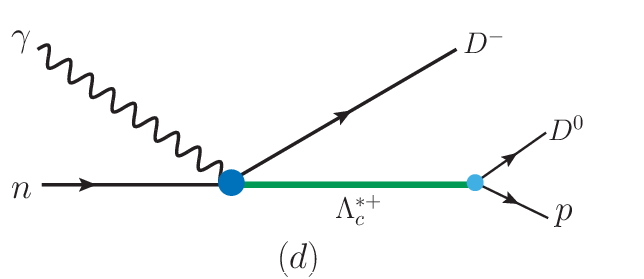}
\end{center}
\caption{(Color online) Feynman diagrams for the $\protect\gamma %
n\rightarrow D^{-}D^{0}p$ reaction.}
\end{figure*}

In Ref. \cite{dong10,dong101}, by assuming the $\Lambda _{c}^{\ast }(2940)$
as a molecular $D^{\ast 0}p$ state, the spin-parity ($J^{P}$) quantum number
of $\Lambda _{c}^{\ast }(2940)$ was assigned to be $\frac{1}{2}^{+}$, while
the quantum number $J^{P}=\frac{1}{2}^{-}$ is completely excluded because
the calculated partial widths are much larger than the experimental width of
$\Lambda _{c}^{\ast }(2940)$ state. In this present work, two cases of $%
\Lambda _{c}^{\ast }(2940)$ with $J^{P}=\frac{1}{2}^{\pm }$ are calculated
for a comparison. Thus we take the normally used effective Lagrangians for $%
\Lambda _{c}^{\ast }ND$, $\Lambda _{c}^{\ast }ND^{\ast }$ and $\gamma
\Lambda _{c}^{\ast }\Lambda _{c}^{\ast }$ couplings as \cite{dong14,xie15},

\begin{eqnarray}
\mathcal{L}_{ND\Lambda _{c}^{\ast }(\frac{1}{2}^{\pm })} &=&ig_{_{\Lambda
_{c}^{\ast }ND}}^{\pm }\bar{\Lambda}_{c}^{\ast }\Gamma ^{\pm }ND+h.c., \\
\mathcal{L}_{ND^{\ast }\Lambda _{c}^{\ast }(\frac{1}{2}^{\pm })}
&=&g_{\Lambda _{c}^{\ast }ND^{\ast }}^{\pm }\bar{\Lambda}_{c}^{\ast }\Gamma
_{\mu }^{\pm }ND_{\mu }^{\ast }+h.c., \\
\mathcal{L}_{\gamma \Lambda _{c}^{\ast }\Lambda _{c}^{\ast }(\frac{1}{2}%
^{\pm })} &=&-e\bar{\Lambda}_{c}^{\ast }(Q_{\Lambda _{c}^{\ast }}%
\rlap{$\slash$}A-\frac{\kappa _{\Lambda _{c}^{\ast }}^{\pm }}{4m_{\Lambda
_{c}^{\ast }}}\sigma ^{\mu \nu }F^{\mu \nu })\Lambda _{c}^{\ast }+h.c.,
\end{eqnarray}%
with
\begin{equation}
\Gamma ^{\pm }=\binom{\gamma _{5}}{1},\text{ }\Gamma _{\mu }^{\pm }=\binom{%
\gamma ^{\mu }}{\gamma _{5}\gamma ^{\mu }}.\text{\ }
\end{equation}%
The $Q_{\Lambda _{c}^{\ast }}$ is the electric charge (in the unite of $e$),
while the anomalous magnetic momentum\footnote{%
In Ref. \cite{af06}, the magnetic moment of lighter state $\Lambda
_{c}(2286) $ is predicted to be 0.38. Since this predicted magnetic moment
does not depend on mass of $\Lambda _{c}$ state, it is reasonable to take $%
\kappa _{\Lambda _{c}^{\ast }}^{+}=0.38$ for the $\Lambda _{c}^{\ast }$ with
$J^{P}=\frac{1}{2}^{+}$.} $\kappa _{\Lambda _{c}^{\ast }}^{+}=0.38$ for the $%
\Lambda _{c}^{\ast }$ with $J^{P}=\frac{1}{2}^{+}$ \cite{af06}. The
anomalous magnetic moment $\kappa _{\Lambda _{c}^{\ast }}^{-}$ for $\Lambda
_{c}^{\ast }$ with $J^{P}=\frac{1}{2}^{-}$ amounts to $0.44$ in the SU(3)
quark model \cite{ra91}. We take the coupling constants $g_{_{\Lambda
_{c}^{\ast }ND}}^{+}=-0.45$, $g_{_{\Lambda _{c}^{\ast }ND}}^{-}=-0.97$, $%
g_{\Lambda _{c}^{\ast }ND^{\ast }}^{+}=6.64$ and $g_{\Lambda _{c}^{\ast
}ND^{\ast }}^{-}=3.75$ as used in Refs. \cite{dong14,xie15}.

Moreover, the effective Lagrangians for the $\gamma DD,\gamma DD^{\ast },$%
and $\gamma NN$ couplings are%
\begin{eqnarray}
\mathcal{L}_{\gamma DD} &=&ieA_{\mu }(D^{+}\partial ^{\mu }D^{-}-\partial
^{\mu }D^{+}D^{-}), \\
\mathcal{L}_{\gamma DD^{\ast }} &=&g_{\gamma DD^{\ast }}\epsilon _{\mu \nu
\alpha \beta }(\partial ^{\mu }A^{\nu })(\partial ^{\alpha }D^{\ast \beta
})D+h.c., \\
\mathcal{L}_{\gamma NN} &=&-e\bar{N}(Q_{N}\rlap{$\slash$}A-\frac{\kappa _{N}%
}{4m_{N}}\sigma ^{\mu \nu }F^{\mu \nu })N,
\end{eqnarray}%
where $F^{\mu \nu }=\partial ^{\mu }A^{\nu }-\partial ^{\nu }A^{\mu }$ with $%
A^{\mu }$, $D$, $D^{\ast \mu }$ and $N$ are the photon, $D$-meson, $D^{\ast
} $-meson and nucleon fields, respectively. $m_{D}$ and $m_{N}$ are the
masses of the $D$-meson and nucleon, while $\epsilon _{\mu \nu \alpha \beta
} $ is the Levi-Civit$\grave{a}$ tensor. $Q_{N}$ is the charge of the hadron
in the unit of $e=\sqrt{4\pi \alpha }$ with $\alpha $ being the
fine-structure constant. The anomalous magnetic moment $\kappa _{N}=-1.913$
for the neutron \cite{yh13}.

The coupling constant $g_{\gamma DD^{\ast }}$ are determined by the
radiative decay widths of $D^{\ast },$%
\begin{equation}
\Gamma _{D^{\ast \pm }\rightarrow D^{\pm }\gamma }=\frac{g_{\gamma DD^{\ast
}}^{2}(m_{D^{\ast }}^{2}-m_{D}^{2})^{2}}{32\pi m_{D^{\ast }}^{2}}\left\vert
\vec{p}_{D}^{~\mathrm{c.m.}}\right\vert ,
\end{equation}%
where $\vec{p}_{D}^{~\mathrm{c.m.}}$ is the three-vector momentum of the $D$
in the $D^{\ast }$ meson rest frame. With $m_{D^{\ast }}=2.01$ GeV, $%
m_{D}=1.87$ GeV and $\Gamma _{D^{\ast \pm }\rightarrow D^{\pm }\gamma }=1.35$
keV, one obtains $g_{\gamma DD^{\ast }}=0.117$ GeV$^{-1}$.

Considering the internal structure of hadrons, a form factor is introduced
to describe the possible off-shell effects in the amplitudes. For the
exchange baryons, we adopt the following form factors as used in Refs. \cite%
{dong14,mosel98,mosel99},%
\begin{equation}
\mathcal{F}_{B}(q_{ex}^{2})=\frac{\Lambda _{B}^{4}}{\Lambda
_{B}^{4}+(q_{ex}^{2}-m_{ex}^{2})^{2}},
\end{equation}%
while for the $D$ and $D^{\ast }$ exchange, we take

\begin{equation}
\mathcal{F}_{D/D^{\ast }}(q_{ex}^{2})=\frac{\Lambda _{D/D^{\ast
}}^{2}-m_{ex}^{2}}{\Lambda _{D/D^{\ast }}^{2}-q_{ex}^{2}}
\end{equation}%
where $q_{ex}$ and $m_{ex}$ are the four-momenta and the mass of the
exchanged hadron, respectively. The values of cutoff parameters $\Lambda
_{B} $ and $\Lambda _{D/D^{\ast }}$ will be discussed in the next subsection.

For the propagators of spin-1/2 baryon, we adopt the Breit-Wigner form \cite%
{dong14,xie15}%
\begin{equation}
G_{1/2}(q_{ex})=i\frac{\rlap{$\slash$}q_{ex}+m_{ex}}{%
q_{ex}^{2}-M_{ex}^{2}+im_{ex}\Gamma }
\end{equation}%
where $\Gamma $ is the total decay width of baryon. We take $\Gamma =17$ MeV
\cite{pdg} for the $\Lambda _{c}^{\ast }(2940)$ state and $\Gamma =0$ for
other intermediate baryons.

The propagator for $D$ exchange is written as%
\begin{equation}
G_{D}(q_{ex})=\frac{i}{q_{ex}^{2}-m_{D}^{2}}
\end{equation}

For the $D^{\ast }$ exchange, we take the propagator as
\begin{equation}
G_{D^{\ast }}^{\mu \nu }(q_{ex})=i\frac{-g^{\mu \nu }+q_{ex}^{\mu
}q_{ex}^{\nu }/m_{D^{\ast }}^{2}}{q_{ex}^{2}-m_{D^{\ast }}^{2}},
\end{equation}%
where $\mu $ and $\nu $ denote the polarization indices of vector meson $%
D^{\ast }.$

\subsection{Cross section for the $\protect\gamma n\rightarrow D^{-}\Lambda
_{c}^{\ast }(2940)^{+}$ reaction}

After the above preparations, the invariant scattering amplitude of $\gamma
(k_{1})n(k_{2})\rightarrow D^{-}(k_{3})\Lambda _{c}^{\ast +}(k_{4})$ process
as shown in Fig. 1 can be constructed as,%
\begin{equation}
-i\mathcal{M}_{j}^{\frac{1}{2}^{\pm }}=\bar{u}(k_{4},\lambda _{\Lambda
_{c}^{\ast }})A_{j}^{\nu (\frac{1}{2}^{\pm })}u(k_{2},\lambda _{n})\epsilon
_{\nu }(k_{1},\lambda _{\gamma }),
\end{equation}%
where $j$ denotes the $s$-, $t$-, $u$-channel or contact term process that
contribute to the total amplitude, while $\epsilon $ and $u$ are the photon
polarization vector and Dirac spinor, respectively. $\lambda _{\Lambda
_{c}^{\ast }}$, $\lambda _{n}$ and $\lambda _{\gamma }$ are the helicities
for the $\Lambda _{c}^{\ast }(2940)$, the neutron, and the photon,
respectively.

The reduced $A_{j}^{v(\frac{1}{2}^{\pm })}$ amplitudes read as%
\begin{eqnarray}
A_{s}^{\nu (\frac{1}{2}^{\pm })} &=&-ie\frac{g_{_{\Lambda _{c}^{\ast
}ND}}^{\pm }}{2m_{N}}\frac{\kappa _{N}}{s-m_{N}^{2}}\Gamma ^{\pm }(%
\rlap{$\slash$}q_{n}+m_{N})\gamma ^{\nu }\rlap{$\slash$}k_{1}\mathcal{F}_{B},
\\
A_{t,D}^{\nu (\frac{1}{2}^{\pm })} &=&-eg_{_{\Lambda _{c}^{\ast }ND}}^{\pm
}\Gamma ^{\pm }\frac{(2k_{3}-k_{1})^{\nu }}{t-m_{D}^{2}}\mathcal{F}_{D}^{2},
\\
A_{t,D^{\ast }}^{\nu (\frac{1}{2}^{\pm })} &=&\frac{g_{\gamma DD^{\ast
}}g_{\Lambda _{c}^{\ast }ND^{\ast }}^{\pm }}{t-m_{D^{\ast }}^{2}}\epsilon
_{\mu \alpha \nu \beta }k_{1}^{\alpha }q_{D^{\ast }}^{\beta }\Gamma _{\mu
}^{\pm }\mathcal{F}_{D^{\ast }}^{2}, \\
A_{u}^{\nu (\frac{1}{2}^{\pm })} &=&-ie\frac{g_{_{\Lambda _{c}^{\ast
}ND}}^{\pm }}{u-m_{\Lambda _{c}^{\ast }}^{2}}\Gamma ^{\pm }[Q_{\Lambda
_{c}^{\ast }}(\rlap{$\slash$}q_{\Lambda _{c}^{\ast }}+m_{\Lambda _{c}^{\ast
}})\gamma ^{\nu }  \notag \\
&&+\frac{\kappa _{\Lambda _{c}^{\ast }}}{2m_{\Lambda _{c}^{\ast }}}(%
\rlap{$\slash$}q_{\Lambda _{c}^{\ast }}+m_{\Lambda _{c}^{\ast }})\gamma
^{\nu }\rlap{$\slash$}k_{1}]\mathcal{F}_{B},
\end{eqnarray}%
where $s=q_{n}^{2}=(k_{1}+k_{2})^{2}\equiv W^{2}$, $t=q_{D/D^{\ast
}}^{2}=(k_{1}-k_{3})^{2},u=q_{\Lambda _{c}^{\ast }}^{2}=(k_{2}-k_{3})^{2}$
are the Mandelstam variables.

To restor the gauge invariance, a generalized contact term is introduced as
\cite{gao10,hh06}%
\begin{equation}
A_{cont.}^{\nu (\frac{1}{2}^{\pm })}=ieg_{_{\Lambda _{c}^{\ast }ND}}^{\pm
}\Gamma ^{\pm }C^{\nu },
\end{equation}%
with%
\begin{eqnarray}
C^{\nu } &=&(2k_{3}-k_{1})^{\nu }\frac{\mathcal{F}_{D}-1}{t-m_{D}^{2}}(1-h(1-%
\mathcal{F}_{B}))  \notag \\
&&+(2k_{4}-k_{1})^{\nu }\frac{\mathcal{F}_{B}-1}{u-m_{\Lambda _{c}^{\ast
}}^{2}}(1-h(1-\mathcal{F}_{D})),
\end{eqnarray}%
where $h=1$ is taken \cite{gao10}.

Thus the unpolarized differential cross section for the $\gamma n\rightarrow
D^{-}\Lambda _{c}^{\ast +}$ reaction at the center of mass (c.m.) frame is
given by
\begin{equation}
\frac{d\sigma }{d\cos \theta }=\frac{1}{32\pi s}\frac{\left\vert \vec{k}%
_{3}^{~\mathrm{c.m.}}\right\vert }{\left\vert \vec{k}_{1}^{{~\mathrm{c.m.}}%
}\right\vert }\left( \frac{1}{4}\sum\limits_{\lambda }\left\vert \mathcal{M}%
\right\vert ^{2}\right)
\end{equation}%
where $\theta $ denotes the angle of the outgoing $D^{-}$ meson relative to
beam direction in the c.m. frame, while $\vec{k}_{1}^{~\mathrm{c.m.}}$ and $%
\vec{k}_{3}^{~\mathrm{c.m.}}$ are the three-momenta of initial $\gamma $ and
final $D^{-}$ meson, respectively.

\subsection{Differential cross section $d\protect\sigma _{\protect\gamma %
n\rightarrow D^{-}D^{0}p}^{2}/dM_{pD^{0}}d\Omega $}

Since the $\Lambda _{c}^{\ast }(2940)$ have a coupling with $pD^{0}$, it is
interesting to discuss the $pD^{0}$ invariant mass or angle distributions
for the Dalitz process $\gamma n\rightarrow D^{-}D^{0}p$. However, it is
difficulty to distinguish the two spin-parity assignments of the $\Lambda
_{c}^{\ast }(2940)$ state from those first order differential cross section
\cite{xie15}. Thus we shall concentrate only on the second order
differential cross section of $d\sigma _{\gamma n\rightarrow
D^{-}D^{0}p}^{2}/dM_{pD^{0}}d\Omega $, which may provide useful information
for clarifying the spin-parity of $\Lambda _{c}^{\ast }(2940)$ state.

The second order differential cross section for the $\gamma n\rightarrow
D^{-}D^{0}p$ reaction\footnote{%
In some theoretical works, it is indicated that the ground state $\Lambda
_{c}(2286)$ also have a coupling with $pD^{0}$. However, it should be noted
that the coupling constant of $\Lambda _{c}(2286)ND$ is determined from $%
SU(4)$ invariant Lagrangians with a great uncertainty. Besides, the mass of $%
\Lambda _{c}(2286)$ is about 650 MeV smaller than that of $\Lambda
_{c}^{\ast }(2940)$, which means that the effects from $\Lambda _{c}(2286)$
state around the $M_{pD^{0}}=m_{\Lambda _{c}^{\ast }}$ should be small
because of the narrow total decay width of $\Lambda _{c}^{\ast }(2940)$
state. Thus the $\Lambda _{c}(2286)$ is not included in this present
calculations.} is written as:

\begin{eqnarray}
\frac{d\sigma _{\gamma n\rightarrow D^{-}D^{0}p}^{2}}{dM_{pD^{0}}d\Omega }
&=&\frac{m_{N}^{2}}{2^{10}\pi ^{5}\sqrt{s}(p_{1}\cdot p_{2})}  \notag \\
&&\int \sum\limits_{spin}\left\vert \mathcal{M}\right\vert ^{2}\left\vert
\vec{p}_{3}\right\vert \left\vert \vec{p}_{5}^{~\mathrm{\ast }}\right\vert
d\Omega _{5}^{\ast },
\end{eqnarray}%
where $M_{pD^{0}}$ is the invariant mass of the final $pD^{0}$ system. $%
\left\vert \vec{p}_{3}\right\vert $ and $\Omega $ are the three-momentum and
solid angle of the final $D^{-}$ meson in the center of mass frame of the
initial $\gamma n$ system, while $\left\vert \vec{p}_{5}^{~\mathrm{\ast }%
}\right\vert $ and $\Omega _{5}^{\ast }$ are the three-momentum and solid
angle of the outing proton in the final $pD^{0}$ system.

\section{Results}

As shown in the previous section, for the $\gamma n\rightarrow D^{-}\Lambda
_{c}^{\ast +}$ process, the $s$-channel with nucleon pole exchange, the $t$%
-channel with $D$ and $D^{\ast }$ exchange as well as the $u$-channel with $%
\Lambda _{c}^{\ast }$ exchange and contact term are considered.
\begin{figure}[tbph]
\begin{center}
\includegraphics[scale=0.4]{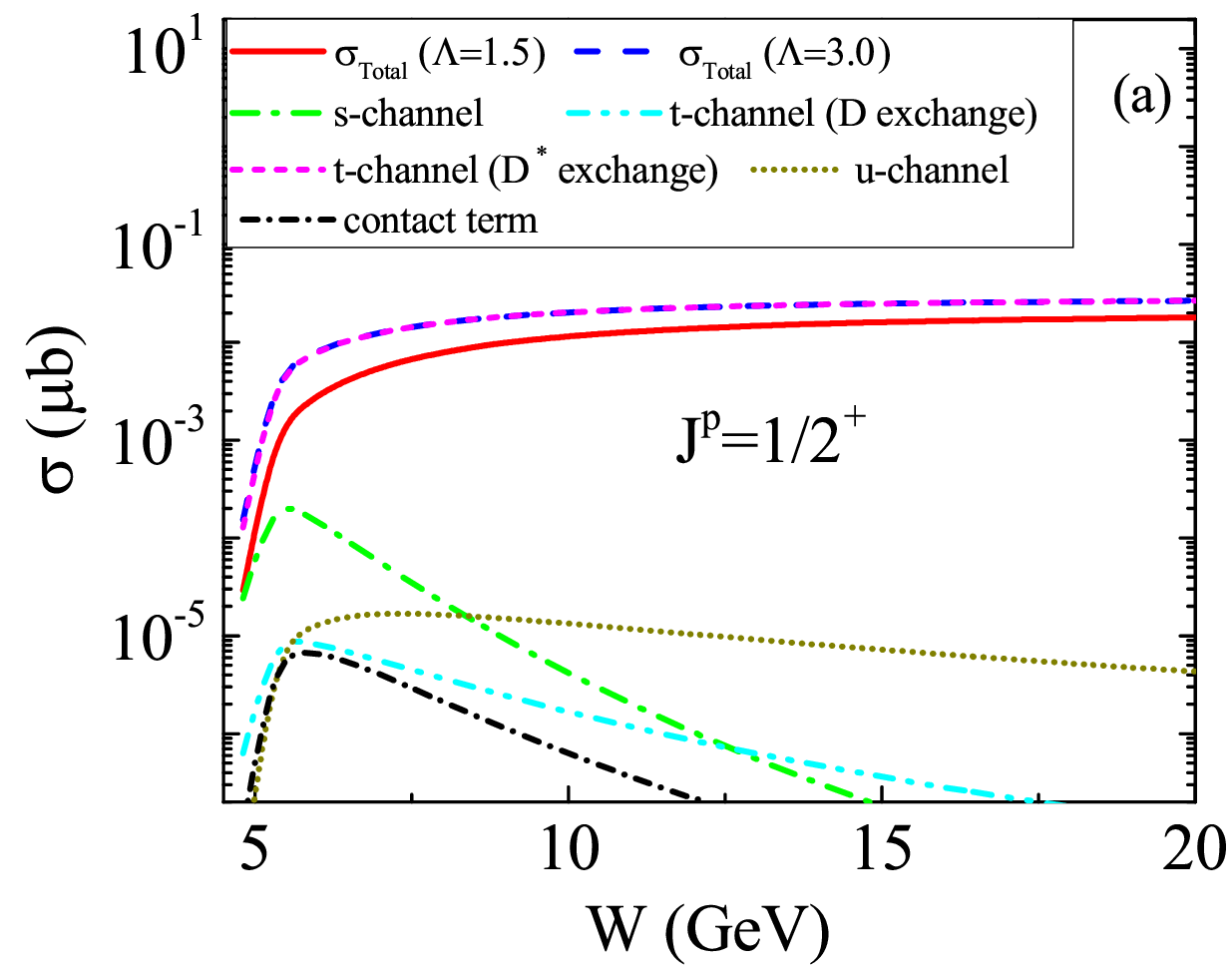} %
\includegraphics[scale=0.4]{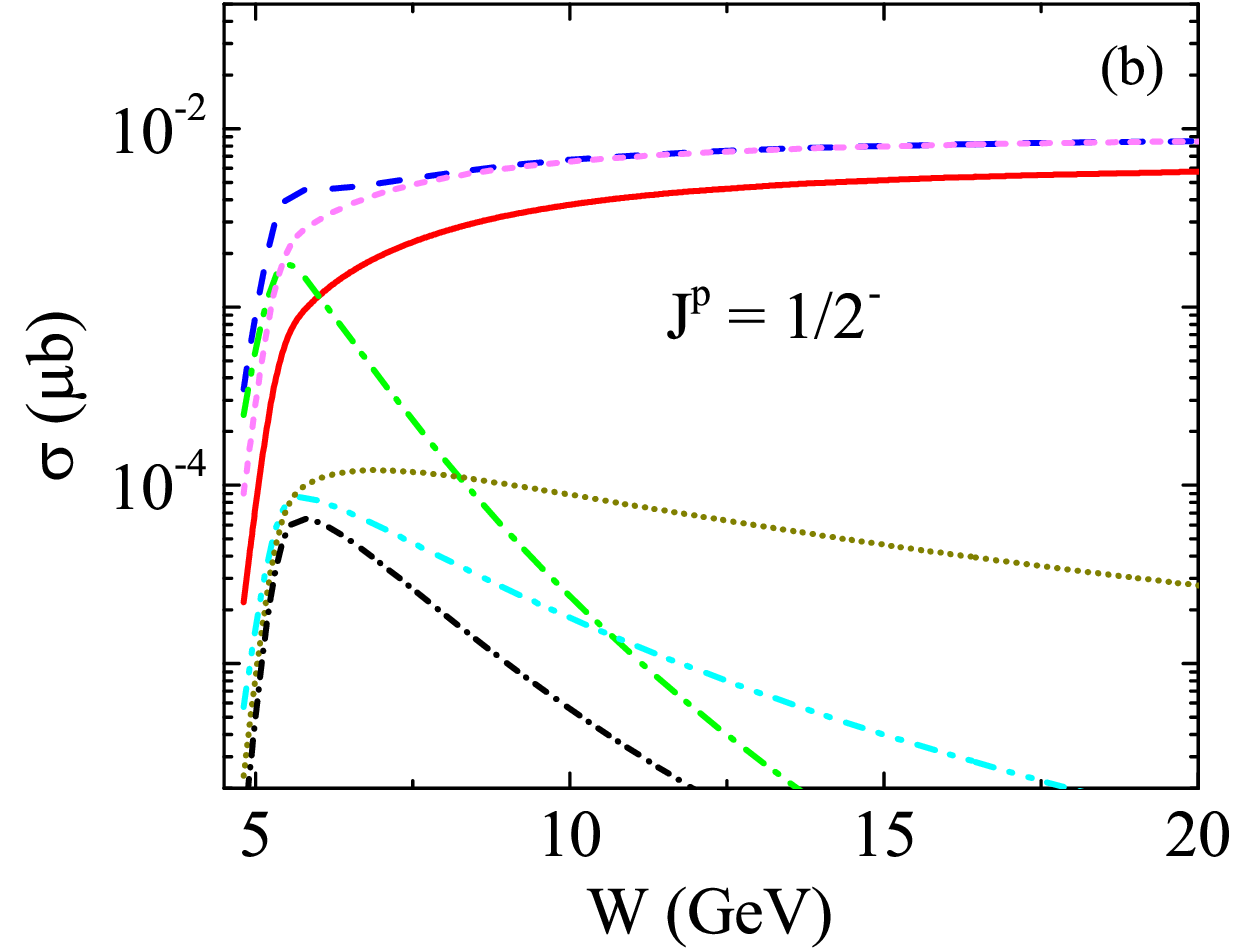}
\end{center}
\caption{(Color online) (a): The total cross section for the $\protect\gamma %
n\rightarrow D^{-}\Lambda _{c}^{\ast +}$ reaction as a function of the
center of mass energy $W$ for the case of $\Lambda _{c}^{\ast }(2940)$ with $%
J^{P}=\frac{1}{2}^{+}$. Here, the $s$-, $t$-, $u$-channel and contact term
are calculated with $\Lambda =3.0$ GeV. (b) is same as the (a), but for the
case of $\Lambda _{c}^{\ast }(2940)$ with $J^{P}=\frac{1}{2}^{-}$.}
\end{figure}

Since the cutoff parameter $\Lambda $ related to the form factor is the only
free parameter, according to usual practice \cite{dong14,xie15,jh10}, we
take the cutoff parameter as $\Lambda =\Lambda _{N}=\Lambda _{D}=\Lambda
_{D^{\ast }}=\Lambda _{\Lambda _{c}^{\ast }}=3.0$ GeV in the sprit of
minimizing the free parameters. For comparison, the numerical results of the
full model with $\Lambda =1.5$ GeV are also presented in Fig. 3, which
indicate the cross section with $\Lambda =1.5$ GeV is smaller than that of $%
\Lambda =3.0$ GeV. Moreover, from fig. 3 one notice that the contribution
from the $t$-channel with $D^{\ast }$ exchange play dominant role\footnote{%
In this work, as mentioned above, the relevant coupling constants are taken
from the Refs. \cite{dong10,dong101} by assuming the charmed $\Lambda
_{c}^{\ast }(2940)$ as a molecular state of $D^{\ast 0}p$. Thus the dominant
$t$-channel with $D^{\ast }$ exchange contribution can be understood easily
since the $\Lambda _{c}^{\ast }(2940)$ have a strong coupling with the $%
D^{\ast 0}p$.} in the $\gamma n\rightarrow D^{-}\Lambda _{c}^{\ast +}$
reaction, while the contribution from the $D$ exchange is very small. The $s$%
-channel with nucleon pole exchange give a considerable contribution near
the threshold. Besides, the contributions from $u$-channel with $\Lambda
_{c}^{\ast }$ exchange and contact term are so small that can be negligible.
With the comparison, it is found that the $s$-channel nucleon pole exchange
have more influence on $\Lambda _{c}^{\ast }(2940)$ with $J^{P}=\frac{1}{2}%
^{-}$ than that of $J^{P}=\frac{1}{2}^{+}$.

Fig. 4 present the differential cross section for $\gamma n\rightarrow
D^{-}\Lambda _{c}^{\ast +}$ process for the cases of $\Lambda _{c}^{\ast
}(2940)$ with $J^{P}=\frac{1}{2}^{\pm }$. It is noticed that All the curves
show strong forward-scattering enhancements, due to the $D^{\ast }$ exchange
in the $t$-channel dominantly.

\begin{figure*}[tbph]
\begin{minipage}{1\textwidth}
\includegraphics[scale=0.38]{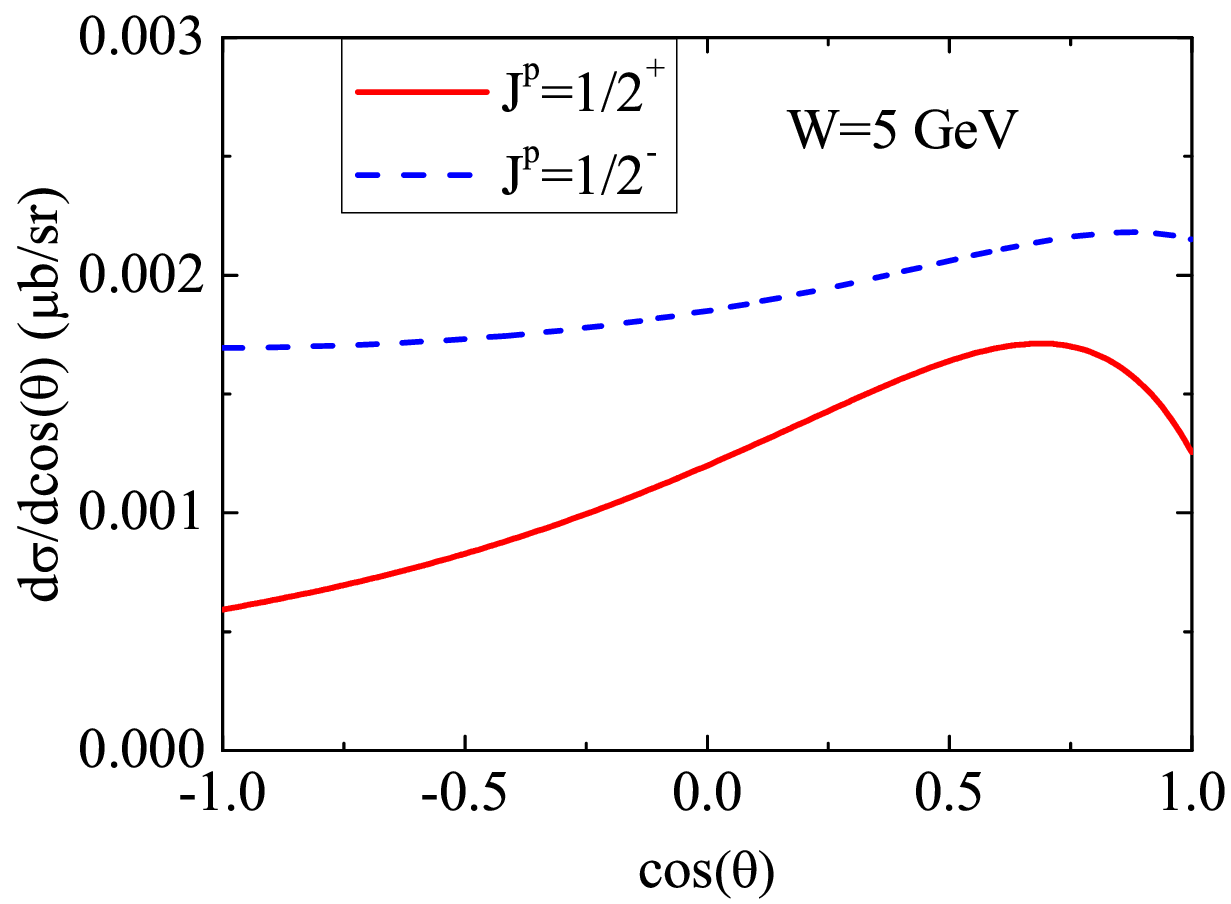}%图形之间的距离
\includegraphics[scale=0.38]{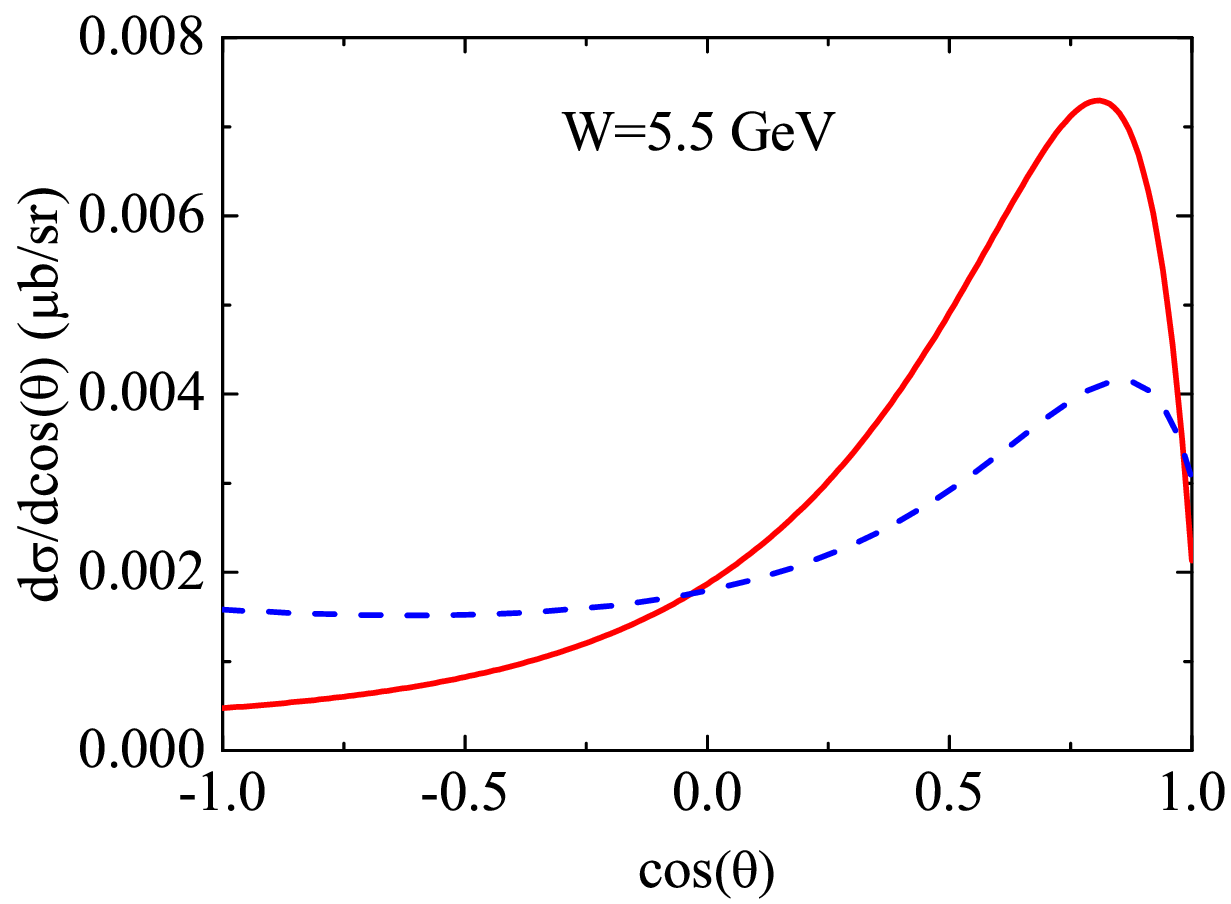}
\includegraphics[scale=0.38]{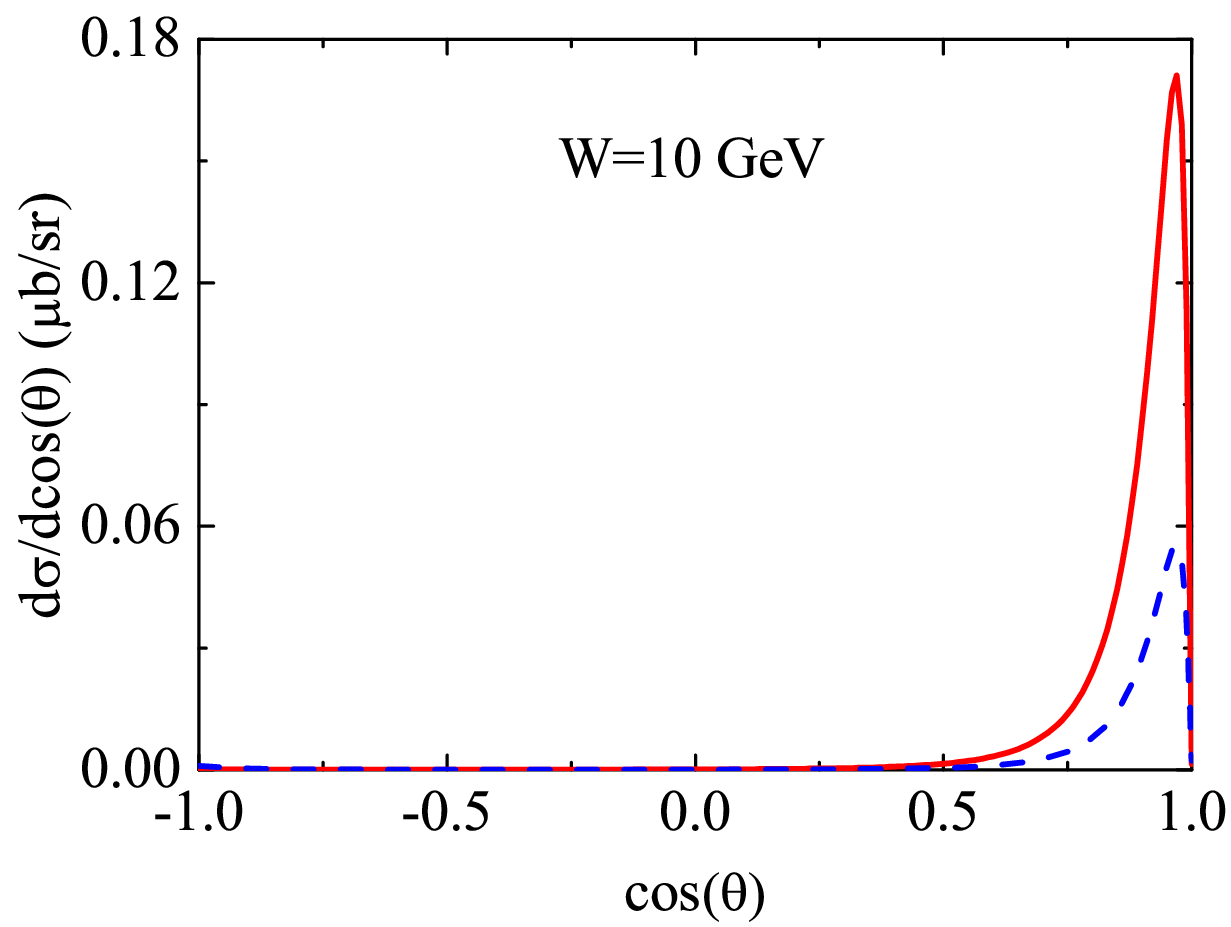}
\includegraphics[scale=0.38]{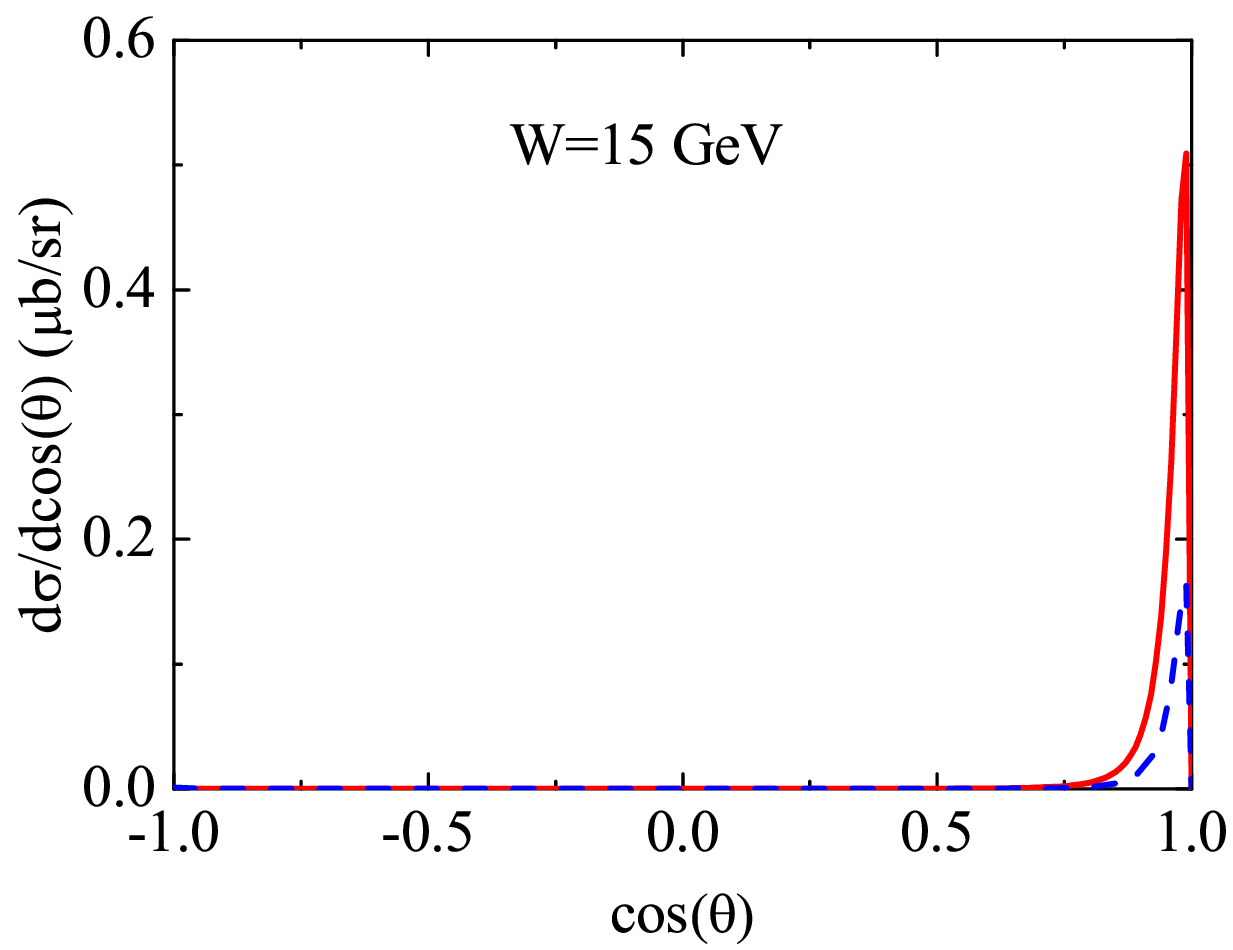}
\end{minipage}
\caption{(Color online) Differential cross section $d\protect\sigma /d\cos
\protect\theta $ as a function of $\cos \protect\theta $ for the $\protect%
\gamma n\rightarrow D^{-}\Lambda _{c}^{\ast +}$ reaction at $W=5,5.5,10,15$
GeV.}
\end{figure*}

Fig. 5 present the differential cross section $d\sigma _{\gamma n\rightarrow
D^{-}D^{0}p}^{2}/dM_{pD^{0}}d\Omega $ at the mass $M_{pD^{0}}=2.94$ GeV for
the cases of $\Lambda _{c}^{\ast }(2940)$ with $J^{P}=\frac{1}{2}^{\pm }$.
It is found that the absolute value of the differential cross section $%
d\sigma _{\gamma n\rightarrow D^{-}D^{0}p}^{2}/dM_{pD^{0}}d\Omega $ for two
spin-parity assignments are much different, which can be checked by further
experiment.
\begin{figure}[tbph]
\begin{center}
\includegraphics[scale=0.4]{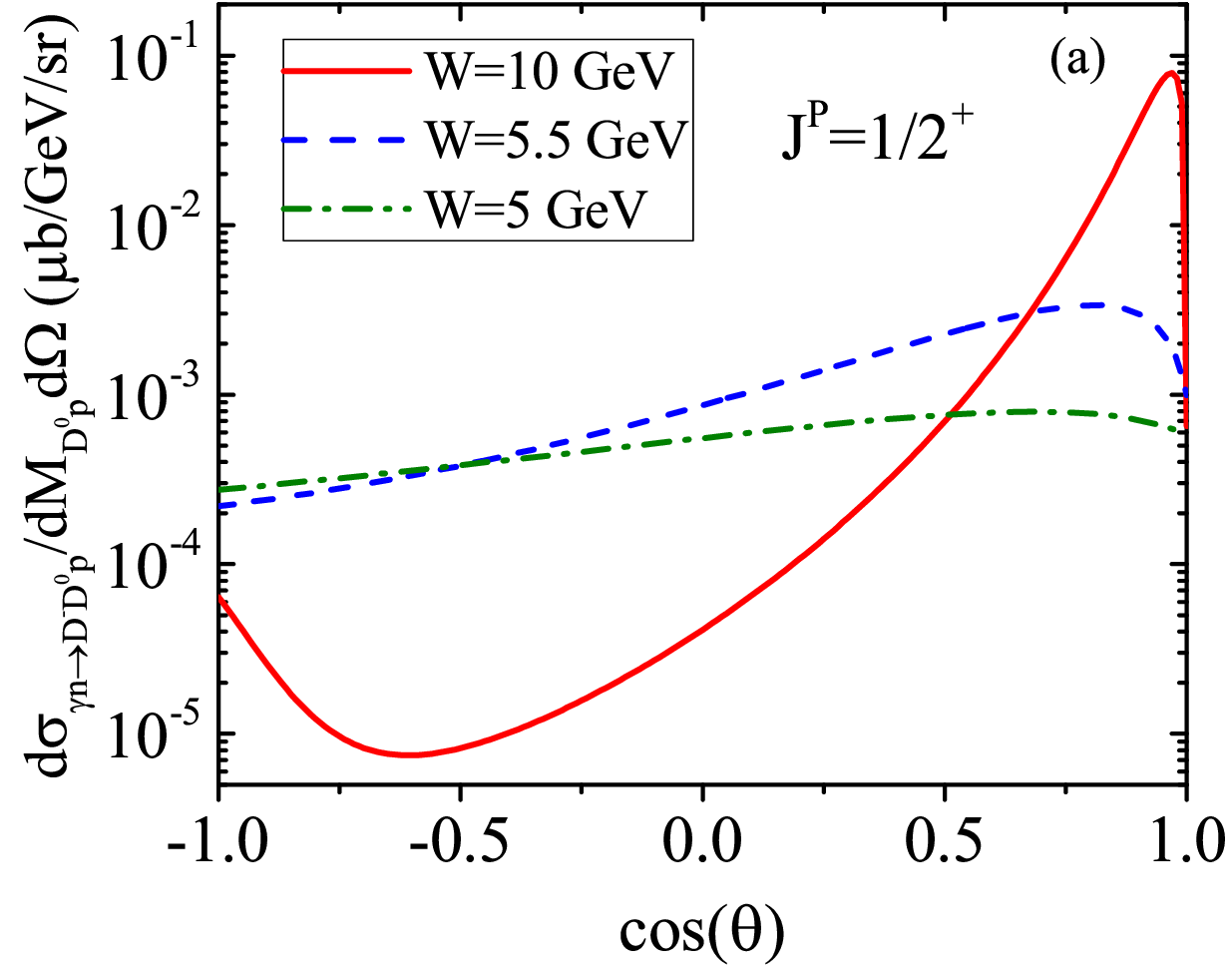} %
\includegraphics[scale=0.4]{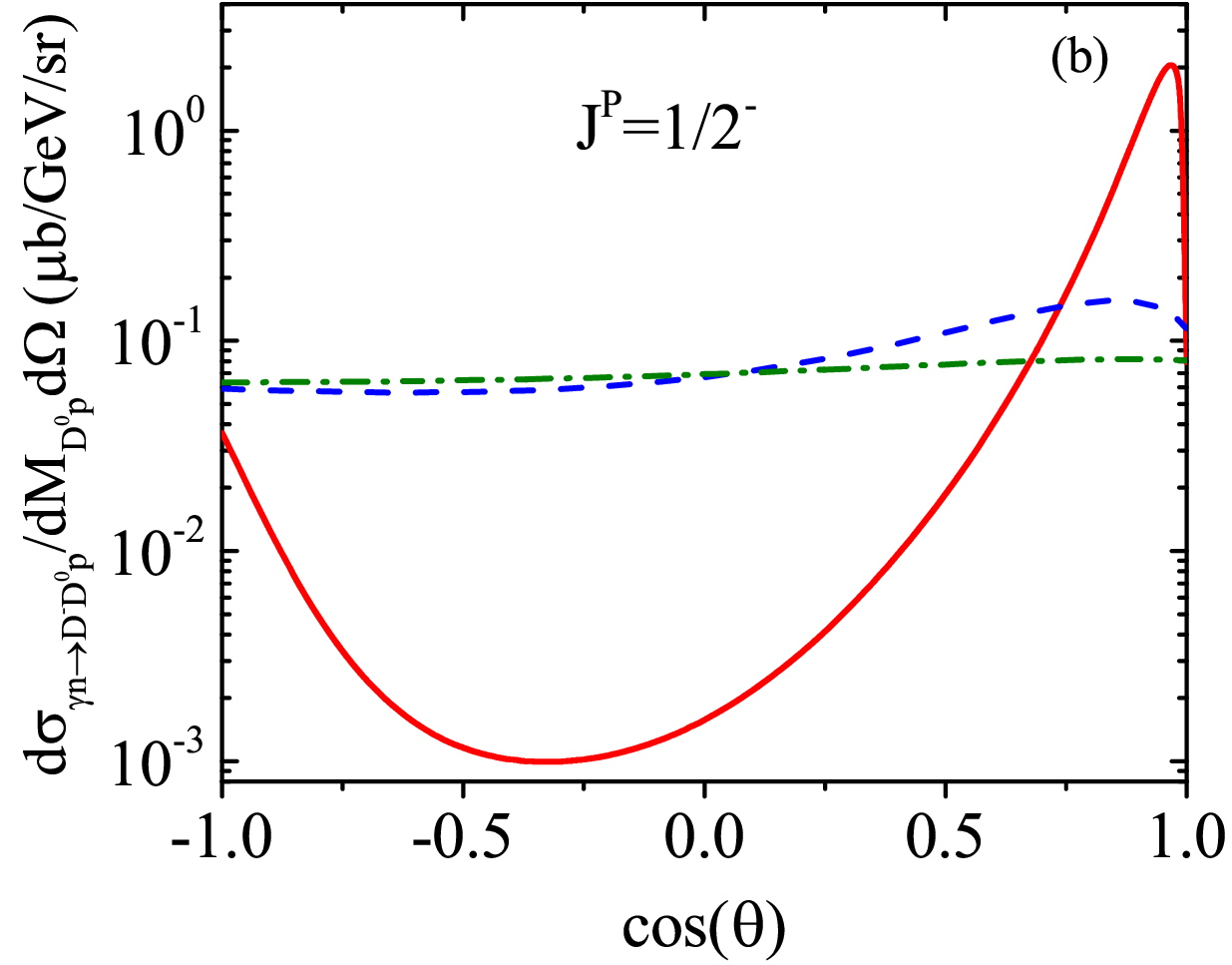}
\end{center}
\caption{(Color online) (a): Differential cross section $d\protect\sigma _{%
\protect\gamma n\rightarrow D^{-}D^{0}p}^{2}/dM_{pD^{0}}d\Omega $ for the
case of $\Lambda _{c}^{\ast }(2940)$ with $J^{P}=\frac{1}{2}^{+}$ at $%
W=5,5.5,10$ GeV. (b) is same as the (a), but for the case of $\Lambda
_{c}^{\ast }(2940)$ with $J^{P}=\frac{1}{2}^{-}$.}
\end{figure}

\section{$\Lambda _{c}^{\ast }(2940)$ production at COMPASS}

The COMPASS experiment at CERN runs since 2002 using positive muon beam of
160~GeV/c (2002-2010) or 200~GeV/c momentum (2011), scattered off solid $%
^{6} $LiD (2002-2004) or NH$_{3}$ targets (2006-2011). It covers the range
of $W$ up to 19.4 GeV. The integrated luminosity of $\gamma N$ interaction
multiplied by the general efficiency of the setup, corresponding the period
of data taking between 2002 and 2011, can be estimated basing on the number
of exclusively produced $J/\psi $ mesons \cite{compass_Zc}. We calculate it
to be of about 10 pb$^{-1}$.

Basing on the integrated luminosity mentioned above and the calculated $%
\Lambda _{c}^{\ast }(2940)$ production cross section value of 0.02 $\mu $b ($%
J^{P}=\frac{1}{2}^{+}$, $\Lambda $=3.0 GeV, $\Gamma _{\Lambda _{c}^{\ast
}\rightarrow pD^{0}}=0.21$ MeV) we can expect to find in the COMPASS muon
data sample collected between 2002 and 2011 up to 0.9$\times 10^{5}$ $%
\Lambda _{c}^{\ast }(2940)$ baryons produced via the reaction $\gamma
n\rightarrow D^{-}\Lambda _{c}^{\ast +}$. This estimation is done neglecting
the nuclear collective effects and assuming the effective amount of neutrons
in the target of about 45\%. This number can be compared with the COMPASS
open charm leptoproduction results based on the data collected between 2002
and 2007 \cite{compass_OCH} where the number of reconstructed $%
D^{0}\rightarrow K^{+}\pi ^{-}$ decays (BR=3.88\%) exceeded $5\times 10^{4}$.

Since the t-channel is dominating, the energy transferred to the produced $%
\Lambda _{c}^{\ast }(2940)$ is small and it decays almost at rest with
momentum of proton and $D^{0}$-meson in the centre-of-mass system of 0.42
GeV/c. Such low-momenta particles are almost invisible for the COMPASS
tracking system while energetic $D^{-}$-meson can be easily detected. So in
spite of impossibility to observe the $\Lambda _{c}^{\ast }(2940)$ decay
directly, its production should manifest itself in the missing mass spectrum.

\section{Summary}

Within the frame of the effective Lagrangian approach, the photoproduction
of charmed $\Lambda _{c}^{\ast }(2940)$ baryon in the $\gamma n\rightarrow
D^{-}\Lambda _{c}^{\ast +}$ process via $s$-, $t$-, $u$-channel and contact
term is investigated based on the conditions of the COMPASS experiment.

The numerical results indicate:

\begin{itemize}
\item[(I)] The $t$-channel with $D^{\ast }$ exchange play dominant role in
the $\gamma n\rightarrow D^{-}\Lambda _{c}^{\ast +}$ reaction, while the
contributions from the $t$-channel $D$ exchange as well the $u$-channel $%
\Lambda _{c}^{\ast }$ exchange and contact term are very small. The $s$%
-channel with nucleon pole exchange give a considerable contribution at the
threshold.

\item[(II)] According to our estimations, a sizable number of events related
to the $\Lambda _{c}^{\ast }(2940)$ {is already} produced at COMPASS
facility, which means it is feasible to searching for the charmed $\Lambda
_{c}^{\ast }(2940)$ baryon produced via $\gamma n$ interaction.{\ In case of
success it would be the first observation of direct production of $\Lambda
_{c}^{\ast }(2940)$.}

\item[(III)] The absolute value of the differential cross section $d\sigma
_{\gamma n\rightarrow D^{-}D^{0}p}^{2}/dM_{pD^{0}}d\Omega $ for the two
assignments $J^{P}=\frac{1}{2}^{\pm }$ for the $\Lambda _{c}^{\ast }(2940)$
state are much different. Thus we suggest this observable can be measured in
the further COMPASS experiment to clarify the nature of $\Lambda _{c}^{\ast
}(2940)$ state.
\end{itemize}

To sum up, we suggest that this experiment be carried out at COMPASS, which
not only helps in testing the above theoretical predictions for the
photoproduction of the $\Lambda _{c}^{\ast }(2940)$ state but also provides
important information for clarifying the nature of the charmed $\Lambda
_{c}^{\ast }(2940)$ baryon. It is worth while pointing out that it is not
possible to give a very precision theoretical result for the production of $%
\Lambda _{c}^{\ast }(2940)$ due to the partial decay width of $\Lambda
_{c}^{\ast }(2940)$ is only a theoretical value but not a real width
measured by experiment. However, from the experimental point of view, the
partial decay width of $\Lambda _{c}^{\ast }(2940)$ is a key factor to
determine the spin-parity of $\Lambda _{c}^{\ast }(2940)$. Thus the
experiment on measuring the partial decay width of $\Lambda _{c}^{\ast
}(2940)$ is also encouraged.

\section{Acknowledgments}

The authors would like to acknowledge Valery Lyubovitskij and Amand Faessler
for useful discussions. Meanwhile, X. Y. W. is grateful Dr. Ju-Jun Xie for
the valuable discussions and help. This project is partially supported by
the National Basic Research Program (973 Program Grant No. 2014CB845406),
the National Natural Science Foundation of China (Grants No. 11175220) and
the the one Hundred Person Project of Chinese Academy of Science (Grant No.
Y101020BR0).

\end{document}